\documentclass[]{aa}
\usepackage[varg]{txfonts}
\usepackage{graphicx,amssymb,amsmath,times,verbatim,pdfpages}
\usepackage[caption=false]{subfig}
\usepackage{natbib}
\bibpunct{(}{)}{;}{a}{}{,}
\usepackage{hyperref}

\begin{document}

\title{Multi-waveband quasi-periodic oscillations in the light curves of blazar CTA 102 during its 2016--2017 optical outburst}

\author{Arkadipta Sarkar\inst{1} \and Pankaj Kushwaha\inst{2}\thanks{Aryabhatta Postdoctoral Fellow} \and Alok C. Gupta\inst{2} \and Varsha R. Chitnis\inst{1} \and Paul J. Wiita\inst{3}}

\offprints{Arkadipta Sarkar, \email{\href{sarkadipta@gmail.com}{sarkadipta@gmail.com}}}

\institute{Department of High Energy Physics, Tata Institute of Fundamental Research, Mumbai 400005, India 
  \and Aryabhatta Research Institute of Observational Sciences (ARIES), Manora Peak, Nainital 263001, India
  \and Department of Physics, The College of New Jersey, PO Box 7718, Ewing, NJ 08628-0718, USA}

\titlerunning{QPO in CTA 102}
\authorrunning{Sarkar et. al.}
\date{Received April 01, 2020 / Accepted August 08, 2020}

\abstract {Quasi-periodic  fluctuations in the light  curves of blazars can  provide insight into the  underlying emission process. This  type of flux modulation  hints at periodic
physical processes that result in  emission. CTA 102, a flat spectrum radio quasar  at a redshift of 1.032, has displayed significant activity  since 2016. The multi-waveband light
curve of  CTA 102 shows signs  of quasi-periodic oscillations during  the 2016--2017 flare.}  {Our goal is to  rigorously quantify the presence  of any possible periodicity  in the
emitted flux  during the mentioned period  and to explore the  possible causes that  can give rise to  it.} {Techniques such as  the Lomb-Scargle periodogram and  weighted wavelet
z-transform were  employed to observe the  power emitted at different  frequencies. To quantify the  significance of the dominant  period, Monte-Carlo techniques were  employed to
consider an underlying smooth bending power-law model for the power spectrum. In addition,  the light curve was modeled using an autoregressive process (AR1) to analytically obtain
the significance  of the  dominant period.  Lastly, the light  curve was  modeled using  a generalized  autoregressive integrated moving  average (ARIMA)  process to  check whether
introducing a seasonal (periodic) component results in a statistically preferable  model.} {Highly significant, simultaneous quasi-periodic oscillations (QPOs) were observed in the
$\gamma$-ray and optical fluxes of blazar CTA  102 during its highest optical activity episode in 2016--2017. The periodic flux modulation had  a dominant period of $\sim$ 7.6 days
and lasted  for $\sim$ 8  cycles (MJD 57710--57770).  All of the  methods used  point toward significant  ($>4\sigma$) quasi-periodic modulation  in both $\gamma$-ray  and optical
fluxes.}{ Several possible models were explored while  probing the origin of the periodicity, and by extension, the 2016--2017 optical  flare. The best explanation for the detected
QPO appears to be a region of enhanced emission (blob), moving helically inside the jet. }

\maketitle

\section{Introduction}
\label{sec:1}
Rapid variability has been one of the identifying criteria of active galactic nuclei (AGNs), which are powered by accretion on the central super massive black hole (SMBH). In
particular the subset of AGNs that are designated as radio-loud have been found to exhibit a diverse range of observational behaviors across all the electromagnetic bands. The
blazar subclass of radio-loud AGNs host powerful large-scale relativistic jets of plasma that is nearly pointed toward us \citep{1995PASP..107..803U}. They emit an entirely
jet-dominated radiation, spread across the entire accessible electromagnetic (EM) spectrum \citep[e.g.,][]{2010ApJ...716...30A}. Blazars are known for their enormous dynamic range
of rarely repeating observational behaviors. In the extragalactic sky, they are the most prominent and persistent broadband nonthermal emitters. Observational studies have
revealed them to be variable on all time scales, all the way from the shortest time scale allowed by the observing facilities to the longest one allowed by the available data
records \citep[e.g.,][and references therein] {2018ApJ...866...11D,2018ApJ...854L..26S}.

In general, accretion-powered sources exhibit variability on a broad range of time scales and exhibit a diverse range of variability behaviors. Despite this, their variability has
been found to share some common statistical variability properties \citep{2015SciA....1E0686S}. Given the fact that accretion-powered sources encompass both compact and noncompact
astrophysical objects of all mass scales from proto-stars to AGNs, with intrinsically very different physical conditions, processes, and mechanisms, the inferred similarity has
been argued to be related to the accretion physics. Blazars' broadband flux variability, despite being dominated by jet emission, also broadly follow these properties
\citep{2016ApJ...822L..13K,2017ApJ...849..138K,2019arXiv191108198B}.

Blazar multi-wavelength (MW) flux variability is, in general, stochastic, but quasi-periodic variations (QPOs) have been reported occasionally \citep[][and references
therein]{1985Natur.314..148V, 1985Natur.314..146C, 1991ApJ...372L..71Q, 1993ApJ...411..614U, 1996A&A...305L..17S, 1996A&A...305...42H, 2003A&A...402..151R, 2008ApJ...679..182E,
2008AJ....135.1384G, 2009ApJ...690..216G, 2019MNRAS.484.5785G, 2009A&A...506L..17L, 2013MNRAS.436L.114K, 2014ApJ...793L...1S, 2016ApJ...820...20S, 2017A&A...600A.132S,
2014JApA...35..307G, 2018Galax...6....1G, 2015ApJ...813L..41A, 2017ApJ...847....7B, 2019MNRAS.487.3990B, 2017ApJ...845...82Z, 2018AJ....155...31H}. The suggested QPO time scales
also have a huge range, similar to the variability time scales shown by blazars, that is, from a few tens of minutes to hours, days, months and even years. \citep[e.g.,][and
references therein]{1968ARA&A...6..417K, 1976ARA&A..14..173S, 1989Natur.337..627M, 1991ApJ...372L...9E, 1993ApJ...404..112S, 1994A&A...288..433B, 1995ApJ...438..120E,
1995ARA&A..33..163W, 1996A&A...305L..17S, 1996A&A...305...42H, 2008AJ....135.1384G, 2012MNRAS.425.3002G, 2015A&A...582A.103G, 2018ApJ...866...11D, galaxies8010015}. In addition to
QPOs, as observed in Galactic black hole X-ray binaries, that are normally attributed to accretion, some QPOs in blazars are also expected from the cosmological hierarchical
structure formation, suggesting the formation of binary supermassive black holes. However, this latter type is expected to be persistent, compared to the accretion induced QPOs
which are generally transient. Further, given the complex physics and not yet fully understood connections between different constituents and jets of AGNs, their imprints, if any,
are expected to be reflected in the jet emission. In this context, the continuous monitoring capability of the Fermi-LAT instrument provides an excellent facility to
explore bright extragalactic $\gamma$-ray sources, most of which are blazars, on both short and long time scales \citep[e.g.,][]{2018ApJ...854L..26S, 2017ApJ...849..138K,
2019MNRAS.484.5785G}.

\begin{figure*}[t]
  \centering
  \subfloat{\includegraphics[width=\linewidth]{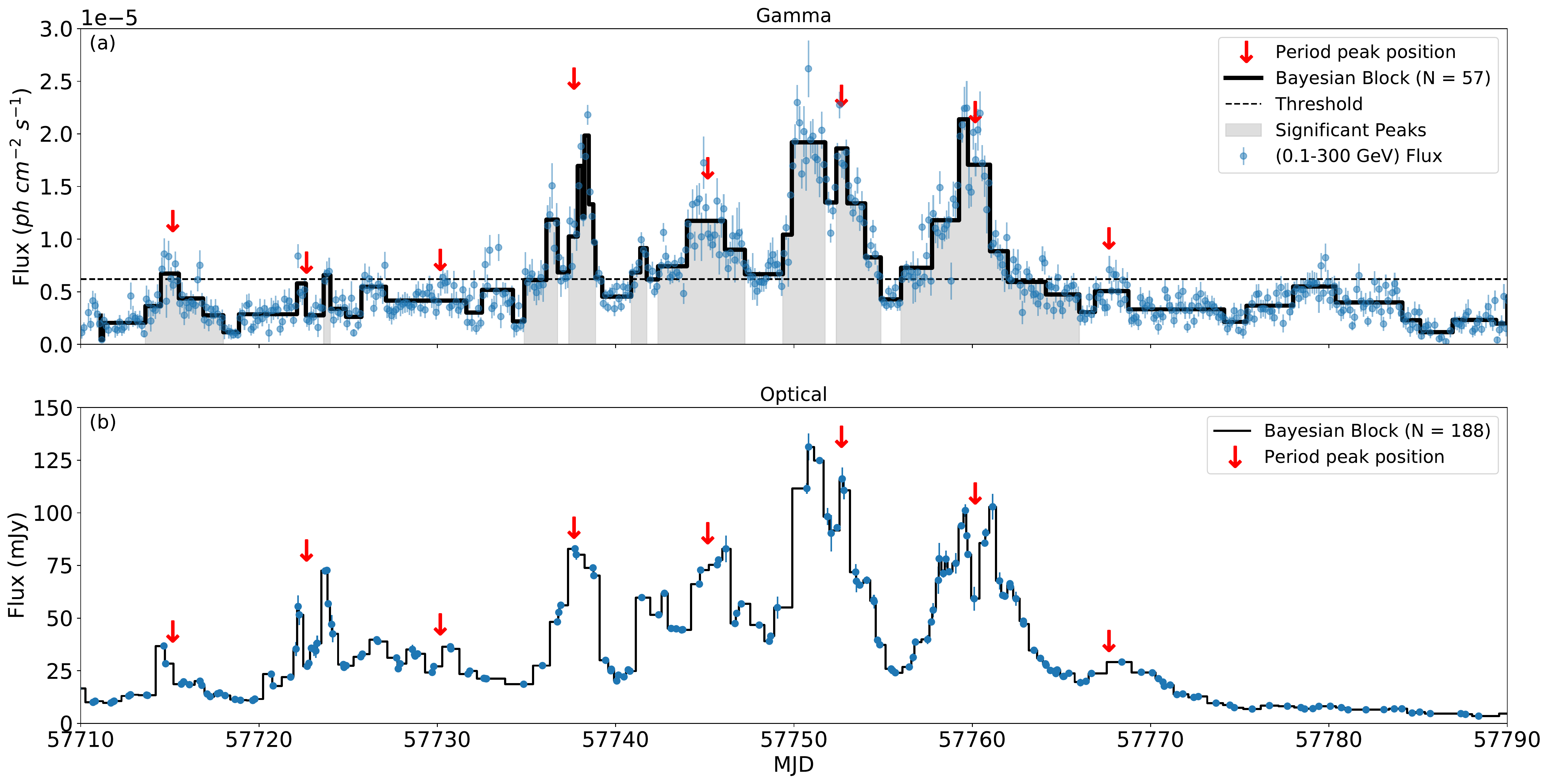}}\\
  \subfloat{\includegraphics[width=\linewidth]{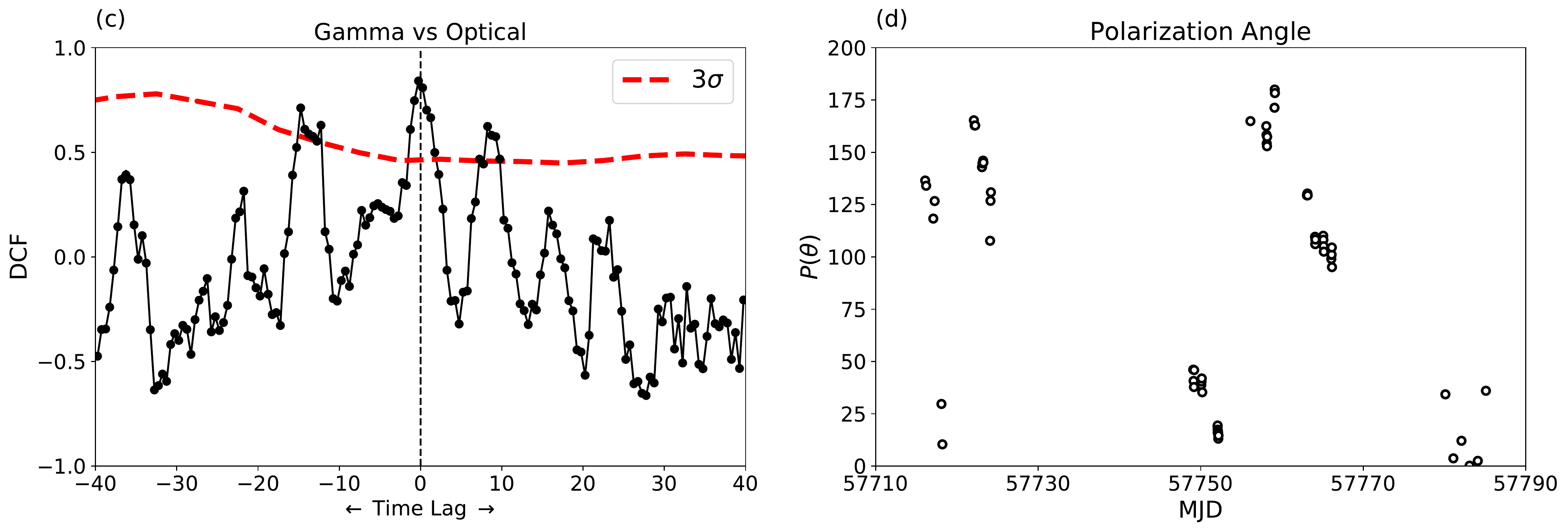}}
\caption{\textbf{CTA 102 light curves during the 2016--17 flare.} \textbf{(a)} Fermi-LAT ($0.1-300$ GeV) light curve. The black line gives the optimal block representation of the
light curve. A peak is considered to be significant if the block height is greater than a set threshold. \textbf{(b)} Optical R band light curve and its optimal block
representation. The QPO peaks are denoted by red arrows. \textbf{(c)} $\gamma$-ray and R-band emission DCF. \textbf{(d)} Optical polarization angle during the detected
QPO. \label{f1}}
\end{figure*}

CTA 102 is a flat spectrum radio quasar (FSRQ) --- the blazar subclass characterized by prominent broad emission lines --- located at the redshift of 1.032
\citep{1965ApJ...141.1295S}. It was first discovered in the radio survey by Owens Valley Radio Observatory (OVRO) at 960 MHz \citep[serial number 102;][]{1960PASP...72..237H} and
its optical counterpart was identified by \citet{1965ApJ...141..328S}. Like other blazars and radio-loud AGNs, it has been explored across multiple EM bands and found to exhibit
the characteristics of radio-loud AGNs, for example, compact star-like appearances in images \citep{1965ApJ...141..328S}, rapid and high broadband variability
\citep{1988AJ.....96.1215P, 2019ApJ...877...39M, 2019MNRAS.490.5300D}, a high ($>3\%$) and variable optical polarization degree \citep{1981ApJ...243...60M, 2015ApJ...813...51C},
high brightness temperature \citep{2013A&A...557A.105F} and nonthermal spectrum \citep[and references therein] {2018ApJ...863..114G, 2017Natur.552..374R}, except that it is more
luminous and has broad emission lines in the optical band only \citep{2016MNRAS.461.3047L}. Optical observations before the year 2000 suggest moderate variability on long terms
\citep{1988AJ.....96.1215P}, but stronger optical variability has been reported after this \citep{2009AJ....138.1902O}. From mid-2011, the source entered an extended high activity
period, exhibiting its highest ever reported flux state from the end of 2016 to the beginning of 2017 \citep[e.g.,][]{2017Natur.552..374R,2019MNRAS.490.5300D}.

\begin{figure*}[t]
\centering
\includegraphics[width=\linewidth]{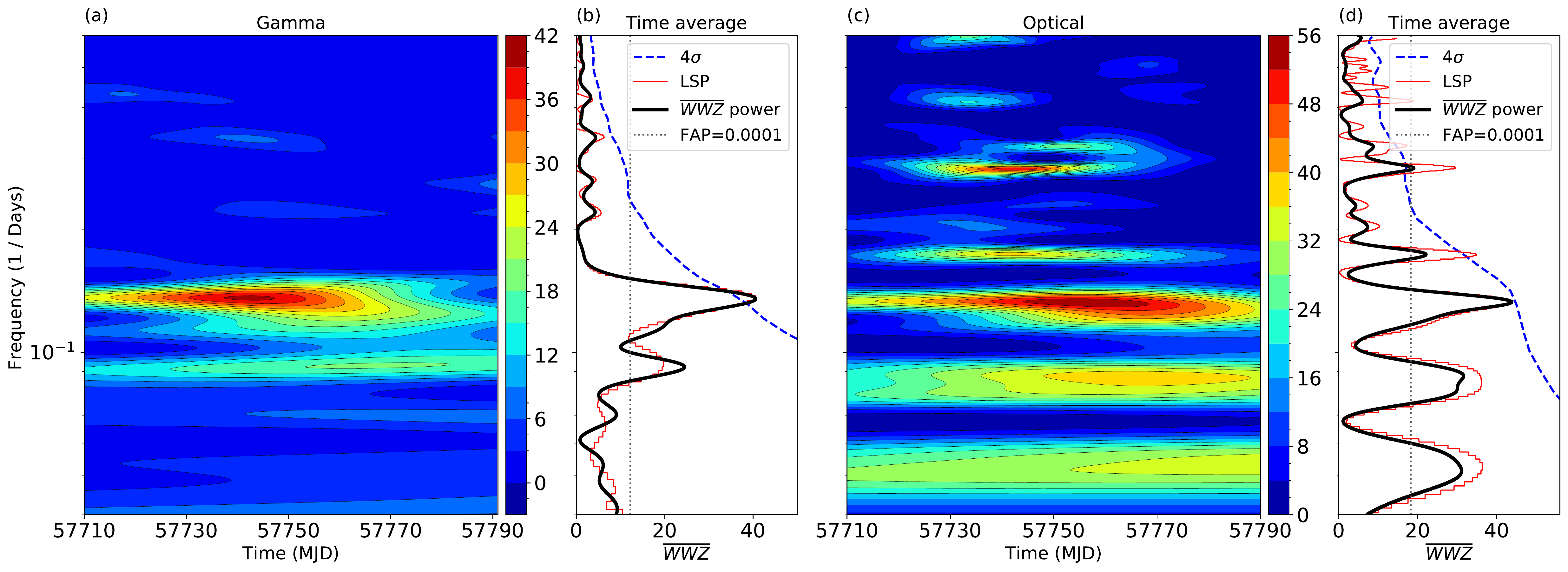}
\caption{\textbf{WWZ maps of CTA 102 during the 2016--17 flare.} \textbf{(a)} $\gamma$-ray WWZ map showing strong periodicity of $\sim 7.6$ days. \textbf{(b)} Time averaged
$\gamma$-ray WWZ map (black) and the LSP (red). The dominant period ($\sim 7.6$ days) shows $> 5\sigma$ significance. \textbf{(c)} Optical R-band WWZ map showing similar
periodicity. \textbf{(d)} Same as \textbf{(b)} for R-band emissions.\label{f2}}
\end{figure*}

At $\gamma$-ray energies, CTA 102 has been detected by the EGRET \citep{1999ApJS..123...79H} and COMPTEL \citep{1995A&A...295..330B} telescopes onboard the Compton Gamma Ray
Observatory (CGRO). It also featured in the first three months of Fermi bright sources list \citep[4C +11.69; ][] {2010ApJ...716...30A}. CTA 102, however, became very active in the
Fermi-LAT band only in 2012, reaching a flux level detectable on daily time scales. Since then its $\gamma$-ray as well as MW activity kept increasing, reaching a peak flux of
$\sim 2\times10^{-5}$~ph~cm$^{-2}$~ s$^{-1}$ in the LAT band at the end of 2017 \citep{2018ApJ...854L..26S, 2019ApJ...871...19Z, 2019ApJ...877...39M, 2019MNRAS.490.5300D}. It was
then the brightest object in the $\gamma$-ray sky with an isotropic luminosity of $\sim 3.25\times10^{50}$~erg ~s$^{-1}$ \citep{2018ApJ...863..114G} on daily time scales. The very
high flux allowed a detailed study of variability down to time scales as small as a few minutes \citep{2018ApJ...854L..26S}. A MW spectral study by \citet{2019ApJ...871...19Z}
argue, that the high activity spread over a few months is due to ablation of a cloud by the relativistic jet and the resulting hadronic interactions. Another possible explanation
is that the high MW activity is likely to involve the interaction between a superluminal component and stationary core $\sim 0.1$ mas from the core, considered to be a
recollimation shock \citep{2019A&A...622A.158C}. On the other hand, a detailed and systematic study by \citet{2017Natur.552..374R} shows that the increased MW activity provides
hardly any evidence of spectral evolution and thus they explained its long term spectral and temporal variability as primarily resulting from variations in the Doppler factor of an
inhomogeneous jet.

In this work, we report the detection of a QPO, in both the $\gamma-$ray and optical R-band light curves of the FSRQ CTA 102, during its highest reported activity period from the end
of 2016 to the beginning of 2017 (MJD: 57710 -- 57790). We found a highly significant QPO of $\sim 7.6$ days. We look at the data reduction methodology in Sect. \ref{sec:2}. The
analysis procedure and the results are in Sects. \ref{sec:3} and \ref{sec:4}. We discuss the possible physical processes explaining the QPO in Sect. \ref{sec:5} and present our
conclusions in Sect. \ref{sec:6}.

\section{Data acquisition}
\label{sec:2}

\subsection{Gamma-ray data:}
\label{sec:2a}
The $\gamma$-ray data used in our work, is taken from the Large Area Telescope (LAT) facility, onboard the Fermi observatory. LAT is an imaging telescope working on the principle
of pair conversion and is sensitive to photon energies $> 20$ MeV. It has a large angular field of view of $2\ sr$ and covers the entire sky every $\sim$ 90 minutes
\citep{2009ApJ...697.1071A}.

We used the PASS8 (\texttt{P8R2}) processed events data of CTA 102 between MJD 57710 -- 57790. The PASS8 data is an improved reconstruction of the entire LAT events and provides a
significant improvement in the data quality \citep{2019arXiv190210045T}. As per the recommendation of the instrument team, we considered only ``SOURCE'' class events
(\texttt{evclass=128, evtype=3}) within the energy range $0.1 - 300$ GeV from a circular region of $15^\circ $ centered on the source. A zenith angle cutoff of $90^\circ$ was
applied during the event selection to avoid contamination from the Earth's limb. Following this, the good time intervals were generated using the standard expression
``\texttt{(DATA\_QUAL>0)\&\&(LAT\_CONFIG==1)}''. Subsequently, an exposure map was calculated with angular coverage $10^\circ$ larger than the region of interest. An XML file
containing the spectral shape of point sources, in the chosen field of view, from the third Fermi-LAT source catalog \citep[3FGL;][]{2015ApJS..218...23A}, was generated.
Additionally, it also included the Galactic and extra-galactic contribution through their respective template files ``\texttt{gll\_iem\_v06}'' and
``\texttt{iso\_P8R2\_SOURCE\_V6\_v06}''. Finally, the optimization over the input XML spectral file was performed using the unbinned likelihood method which comes bundled with the
LAT analysis software.

The $\gamma$-ray light curve of the source was extracted following the above-mentioned procedure for each time interval. First, we extracted the daily light curve of the source by
iteratively removing nonsignificant sources during the likelihood fit as described in \cite{2014ApJ...796...61K} until the analysis converged. We then used the daily best-fit
source XML files to extract the 3h light curve. Following the data reduction techniques mentioned above, the $\gamma$-ray analysis was performed in sequential time steps, with time
bins of 3 hours to obtain the light curve of the source during its flaring state. Bins of three-hour length, optimized both cadence and coverage for the purpose of our analysis;
they were long enough to give adequate signal-to-noise ratios (test statistics > 9) in nearly all bins, and short enough to give a large enough number of points for good light
curve modeling. When rebinned to 24 hour segments, our light curve was in very good agreement with the Fermi-LAT collaboration daily quick-look product. As discussed below, we
checked that the key results were essentially identical when somewhat longer bins were considered.

\subsection{Optical (R-band) data:}
\label{sec:2b}
The \(R\) passband optical photometric data presented here is collected from the various telescopes around the globe, under the whole earth blazar telescope (WEBT) observational
monitoring campaigns of the blazar CTA 102. The original \(R\) passband optical photometric data of these observational campaigns have been published \citep{2017Natur.552..374R,
2019MNRAS.490.5300D}, and were kindly provided to us by C.~M.~Raiteri. The details concerning the \(R\)-band optical photometric data and analysis methods have been published
\citep{2017Natur.552..374R, 2019MNRAS.490.5300D}. We received the data in magnitude vs time format which we converted to flux vs time. For consistency, the optical fluxes were also
binned into 3-hour segments.

\section{Analysis}
\label{sec:3}

The initial hint of quasi periodicity came from the visual inspection of the optical and \(\gamma\)-ray light curve of CTA 102, during its 2016-17 flare (Figure \ref{f1}). We
employed several independent techniques to identify and quantify periodicity. The individual techniques are elaborated below.

\subsection{Light curve inspection}
\label{sec:3a}

Visual inspection of the light curves hinted at quasi-periodicity, owing to several apparently equispaced features (peaks) in the light curve. However, a na\"{i}ve visual inspection
is often subject to bias, hence more rigorous methods are necessary to analyze features in the light curve. To properly define the features, we use the Bayesian Block (BB)
representation of the light curve \citep{2013ApJ...764..167S}. This involves segregating the light curve into its optimal block representation, by maximizing the fitness function
among all segmentations of the light curve. This assumes the photon number distribution to be Poissonian and the False Alarm Probability (FAP) of the representation was \(<0.01\).
The block representation of the light curve helps in identifying the location and extent of significant peaks in the light curve. We define a peak as, a block (\(B\)) where the
block height (average flux) is greater than both the preceding and succeeding blocks. We also obtain the extent of the peak by traveling downward from the peak in both forward and
backward directions, and the peak flux (\(\mathcal{F}_{p}\)) from the height of the block \(B\). By this definition, peaks can occur even in a relatively low flux state. Thus, we
only statistically associate a peak with a flare, if \(\mathcal{F}_{p}\) is greater than \(2 \sigma\) from the mean of the global light curve. \cite{2019ApJ...877...39M} gives the
global properties of the \(\gamma\)-ray light curve and further details on the peak finding method. For the present analysis, we consider a \(\gamma\)-ray peak to be statistically
significant, if \(\mathcal{F}_{p} > 6.6\times10^{-6}\) ph\ cm$^{-2}$\ s$^{-1}$. After identifying the significant peaks in the light curves, the relative separation between the
peaks can help detect quasi-periodicity.

This analysis methodology for peaks does not trivially translate to the \(R\)-band data, since the long-term properties of this light curve (mean and variance) are unknown, even
though some estimates are available at earlier times, for the \(B\) and \(V\) bands \citep{2010A&A...518A..10V} as well as the \(G\), \(H\) and \(K\) bands
\citep{2003yCat.2246....0C}. To alleviate this issue, we used a Discrete Correlation Function \citep[DCF;][]{1988ApJ...333..646E} to quantify the correlation between the
\(\gamma\)-ray and \(R\)-band optical emission. The presence of a significant correlation at zero lag would indicate quasi-periodicity in the \(R\)-band emission if the
\(\gamma\)-ray emission shows a QPO (or vice versa). To calculate the DCF, we first subtract a linear baseline \citep{Welsh1999} and then compute the Unbinned DCF (UDCF) by
\begin{equation}
  \label{eq1}
U D C F_{i j}=\frac{\left(\mathcal{F}_{\gamma}\left(t_{i}\right)-\overline{\mathcal{F}_{\gamma}}\right)\left(\mathcal{F}_{R}\left(t_{j}\right)-\overline{\mathcal{F}_{R}}\right)}{\sqrt{\sigma^{2}\left(\mathcal{F}_{\gamma}\right) \sigma^{2}\left(\mathcal{F}_{R}\right)}},
\end{equation}
here \({\mathcal{F}_A(t_i)}\) is the \(A\) band flux at time \(t_i\) and \(\overline{\mathcal{F}}\) and \(\sigma^2(\mathcal{F})\) are its mean and variance, respectively. We
obtained the DCF by binning the points in the range \(\tau -\Delta \tau/2 \leq \Delta t_{ij} \leq \tau +\Delta \tau /2\). Here \(\tau\) is the time lag, \(\Delta \tau\) is the bin
width and \(\Delta t_{ij} = t_i-t_j\), giving
\begin{equation}
  \label{eq2}
\begin{split}
D C F(\tau) & =\frac{1}{n} \sum_{k=1}^n U D C F_{k}(\tau) \\
& \pm \frac{1}{n-1} \sqrt{\sum_{k=1}^{n}\left(U D C F_{k}-D C F(\tau)\right)^{2}},
\end{split}
\end{equation}
where \(n\) is the number of points binned following the binning criteria. Since blazar emissions are variable, we calculate the means and variances in Eq. \ref{eq1} using only the
points falling within given lag bins \citep{WhitePeterson1994}.

\subsection{Power spectral analysis}
\label{sec:3b}

Light curve inspection, including BB modeling, can only provide an idea about periodicity in a light curve. Proper quantification of a possible QPO is necessary for making any
claim regarding the emission mechanisms at play. Power spectral density (PSD) or periodograms were used to compute the power emitted at different frequencies. PSDs are mod-squares
of the Discrete Fourier transform (DFT) of the source light curve. If emission power is significantly higher in a particular frequency, then we have evidence that there may be a
periodic component to the emission. The DFT algorithm needs uniformly sampled data points (flux in the present case), which is very unlikely for any astrophysical source. We
instead use a Lomb-Scargle Periodogram \citep[LSP,][]{1976Ap&SS..39..447L, 1982ApJ...263..835S}, to take into account irregular sampling. This involves, fitting the light curve
with sinusoidal functions, with different frequencies, and constructing a periodogram from the goodness of the fit. This method can directly detect persistent periodicities;
however (as in this work) astrophysical periodicities could be transient. Transient periodicities could arise from effects in the blazar jet or accretion disk that are short-lived.
A LSP is insensitive to transient periodicities unless we already know the extent of the periodicity stretch a priori.

To detect and quantify any transient QPOs, we used the Weighted Wavelet Z-transform method \citep[WWZ,][]{1996AJ....112.1709F}. This method decomposes the data into time and
frequency domains (WWZ maps), by convolving the light curve with a time and frequency-dependent kernel. The present analysis uses the Morlet Kernel \citep{doi:10.1137/0515056} with
analytical form \(f(\omega[t-\tau])=\exp(i \omega(t-\tau)-c \omega^{2}(t-\tau)^{2})\). The WWZ map is then given by:
\begin{equation}
  \label{eq3}
\begin{split}
  W(\omega, \tau ; x(t))&=\omega^{1 / 2} \int x(t) f^{*}(\omega(t-\tau)) d t \\
  &=\omega^{-1 / 2} \int x\left(\omega^{-1} z+\tau\right) f^{*}(z) d z .
\end{split}
\end{equation}
Here \(f^*\) is the complex conjugate of the wavelet kernel \(f\), \(\omega\) is the scale factor (frequency), \(\tau\) is the time-shift, and \(z=\omega(t-\tau)\). The kernel acts
as a windowed DFT, with the window \(\exp(-c \omega^{2}(t-\tau)^{2})\), with the window size depending on \(\omega\) and the constant \(c\). In the present analysis, we obtain the
best value of the parameter \(c\) by matching the time-averaged WWZ and the LSP in the same time period. A WWZ map has the advantage of locating both, any dominant periods and
their spans in time and hence is useful in locating the transient periodicities.

\subsection{Light curve modeling}
\label{sec:3c}

Blazar light curves (like any time-series) can be modeled using stochastic models. Modeling the light curve can reveal periodicity in the emission process. When the present
emission depends on the past emissions (yielding fewer drastic jumps in the emission), the model used is an Autoregressive (AR) model. It is mathematically expressed as \(
\mathcal{F}(t_{i}) = \sum_{j=1}^p \theta_j\ \mathcal{F}(t_{i-j}) + \epsilon (t_{i})\), where \(\mathcal{F}(t_i)\) is the emission at time \(t_i\) which depends on the \(p\) prior
emissions, with \(\theta_i\) the auto-regression coefficients, and \(\epsilon(t)\) is a normally distributed random variable (fluctuation or forecasting error). Similarly, when the
present emission depends on the past fluctuations, the model used is a Moving Average (MA) model, where \( \mathcal{F}(t_{i}) = \sum_{j=1}^q \phi_j \epsilon(t_{i-j}) + \epsilon
(t_{i})\) and as before, \(\epsilon(t_i)\) are normally distributed random variables.

The combined Autoregressive Moving Average (ARMA) model can nicely explain stationary timeseries; however, blazar light curves are often nonstationary. To deal with this issue, it
is helpful to transform the light curve by successive differencing (\(\Delta^d\)), defined as: \(\Delta\mathcal{F}(t_i) = \mathcal{F}(t_i) -\mathcal{F}(t_{i-1})\), prior to
modeling. This more complex model is then termed an Autoregressive (AR) Integrated (I) Moving Average (MA) or ARIMA\((p, d, q)\) model \citep{1981ApJS...45....1S,
2018FrP.....6...80F}, where the ``integrated'' part in the model accounts for the successive differencing. It is given by
\begin{equation}
  \label{eq6}
  \begin{split}
    &\Delta^d \mathcal{F}(t_{i}) = \sum_{j=1}^p \theta_j\ \Delta^d\mathcal{F}(t_{i-j}) + \sum_{j=1}^q \phi_j \epsilon(t_{i-j}) + \epsilon(t_i)\\
    &\text{or,}\ \left(1 - \sum_{j=1}^p \theta_j \mathcal{L}^j \right)\Delta^d \mathcal{F}(t_i) = \left(1 - \sum_{j=1}^q \phi_j \mathcal{L}^j \right) \epsilon(t_i) ,
  \end{split}
\end{equation}
where the successive differencing operator is defined as: \(\Delta^d = (1-\mathcal{L})^d\). In the above equation, the second representation of the ARIMA model uses a lag operator
defined as \(\mathcal{L}^k\mathcal{F}(t_i)=\mathcal{F}(t_{i-k})\) and \(\mathcal{L}^k \epsilon(t_i) = \epsilon(t_{i-k})\) and \(p\), \(q\) and \(d\) are the order of AR, MA, and
differencing respectively.

An ARIMA(\(p,d,q\)) approach can sufficiently model any nonperiodic light curve. However, in the presence of periodicity, it is preferable to use a Seasonal (S) Autoregressive
(AR) Integrated (I) Moving Average (MA) or SARIMA\((p,d,q)\times(P,D,Q)_s\) model \citep{Adhikari2013AnIS}, which can be expressed as
\begin{equation}
  \label{7}
  \begin{split}
  \implies  &\left( 1 - \sum_{j=1}^p \theta_j \mathcal{L}^j \right) \left( 1 - \sum_{j=1}^{P} \Theta_j \mathcal{L}^{sj} \right) \Delta^d\Delta_s^D\mathcal{F}(t_i) \\
   = &\left( 1 + \sum_{j=1}^p \phi_j \mathcal{L}^j \right) \left( 1 + \sum_{j=1}^{Q} \Phi_j \mathcal{L}^{sj} \right) \epsilon(t_i),
  \end{split}
\end{equation}
where the second factor in both LHS and RHS is the seasonal term responsible for periodicity with a period \(s\), \(\Delta^D\mathcal{F}(t_i)= (1-\mathcal{L}^s)^D \mathcal{F}(t_i)\)
is the order of seasonal differencing, and \(P\) and \(Q\) are the orders of seasonal AR and MA, respectively.

Then the search for periodicity involves fitting the light curve using both ARIMA and SARIMA models and comparing their goodness-of-fit. We used the Akaike information criterion
\citep[AIC,][]{1100705} for model comparisons. The AIC\(\equiv -2lnL+2k\), where \(L\) is the likelihood of obtaining the data given the model and \(k\) is the number of free
parameters in the model. AIC rewards a model for fitting the data better while penalizing it for using a larger number of parameters. We favor the model with the lowest AIC during
model comparison. In the present case, if there is no periodicity, models with \((p,d,q)\times(0,0,0)_s\) will have a lower AIC. We performed a grid search of AIC values in the
parameter space
\begin{equation}
  \label{eq8}
\phi=
    \begin{cases}
      p,q &\in [0,10]\\
      P,Q &\in [0,6]\\
      d,D &\in \{0,1\}\\
      s &\in [0,10]\ \text{days}
    \end{cases}
\end{equation}
and obtained the most likely model for the light curve. If a model with a periodic component better explains the light curve, the lowest AIC value will be in a model with nonzero
$P,\ Q,\ \text{or}\ D$.

\subsection{Significance estimation}
\label{sec:3d}

Along with constructing the PSDs and identifying any dominant period, it is necessary to determine its significance. The significance of the PSD peak quantifies how likely it is to
obtain a particular peak power due to random fluctuations, given the underlying model. This requires an assumption for the underlying model of the periodogram and, by extension, the
underlying model of the light curve. The most basic model, ARIMA(\(0,0,0\)), assumes that each emission is independent, \(\mathcal{F}(t_i) = \epsilon(t_i)\), which produces a PSD
where the emitted power in different frequencies are independent (white-noise). We can model such a PSD using a constant in the frequency domain. With \(N\) different PSD
frequencies, the probability (\(p\)) that the maximum power at any frequency crossing a threshold (\(z\)) is given by \(p(>z) \approx N\ e^{-z}\) \citep{1982ApJ...263..835S,
2018AJ....155...31H}, quantifying the false alarm probability (FAP), or the significance of the peak. The lower the FAP of a period peak, the less likely it is to be caused by
statistical fluctuation.

We empirically know that the underlying power spectrum of blazars (like most autoregressive process) demonstrates a red-noise periodogram model \citep[e.g.,][]{2005A&A...431..391V},
with more power being emitted at lower frequencies. So using an underlying ARIMA(\(0,0,0\)) or a white-noise model would give wrong estimates for the significance of the peaks.
Ideally it makes sense to use the PSD model corresponding to the (S)ARIMA(\(p,d,q\)) model that best explains the given light curve. However, an analytical form for the periodogram
is available only for ARIMA(\(1,0,0\)) model \citep[or AR1 light curves, ][]{ROBINSON19779}, where the flux at a particular time depends only on the flux that preceded it.
\cite{percival_walden_1993} models the ARIMA(\(1,0,0\)) PSD as
\begin{equation}
  \label{eq5}
  G_{rr}(f_j)=G_0\frac{1-\theta^2}{1-2\theta \cos (\pi f_i/f_{Nyq}) + \theta^2},
\end{equation}
where $f_j$s are the discrete frequencies up to the Nyquist frequency ($f_{Nyq}$), $G_0$ is the average spectral amplitude, $\theta\equiv \exp(\Delta t/\tau)$ is the average
autocorrelation coefficient and \(\Delta t\) is the average sampling interval. We obtain the characteristic autocorrelation timescale ($\tau$) from
Welch-overlapped-segment-averaging \citep[WOSA,][]{1161901} of the LSP. We then estimated the significance of a peak by considering a $\chi^2$ distribution of periodogram values
about the theoretical model. This procedure is performed by the computer code
\citep[REDFIT\footnote{\href{https://www.manfredmudelsee.com/soft/redfit/index.htm}{https://www.manfredmudelsee.com/soft/redfit/index.htm}},][]{Schulz:2002:RER:607225.607238}. The
distribution is $\chi^2$ since the periodogram points are constructed from mod-squaring the real and imaginary parts which are assumed to be normally distributed.

Simpler light curves models, such as pure power-laws \citep[$P\propto \nu^{-\alpha_1}$, ][]{2005A&A...431..391V} or smooth bending power-laws \citep[][]{2010MNRAS.402..307V} could
reasonably approximate the underlying red-noise PSD. We observed (again using AIC) that the underlying model is closer to a smooth bending power-law than a simple power-law. So we
used a smooth bending power-law to model the PSD for non-AR1 light curves. We used Monte Carlo (MC) techniques to simulate one thousand light curves with the same underlying PSD
model and distribution of fluxes (PDF) as the original light curve \citep{2013MNRAS.433..907E}. From the LSP and WWZ analyses of the simulated light curves, we calculated the
significance of the peak from the fraction of simulated light curves where the power crosses that of the original at the dominant frequency. We modeled the underlying PDF with a
log-normal distribution.

\begin{figure}[t]
\centering
\includegraphics[width=\linewidth]{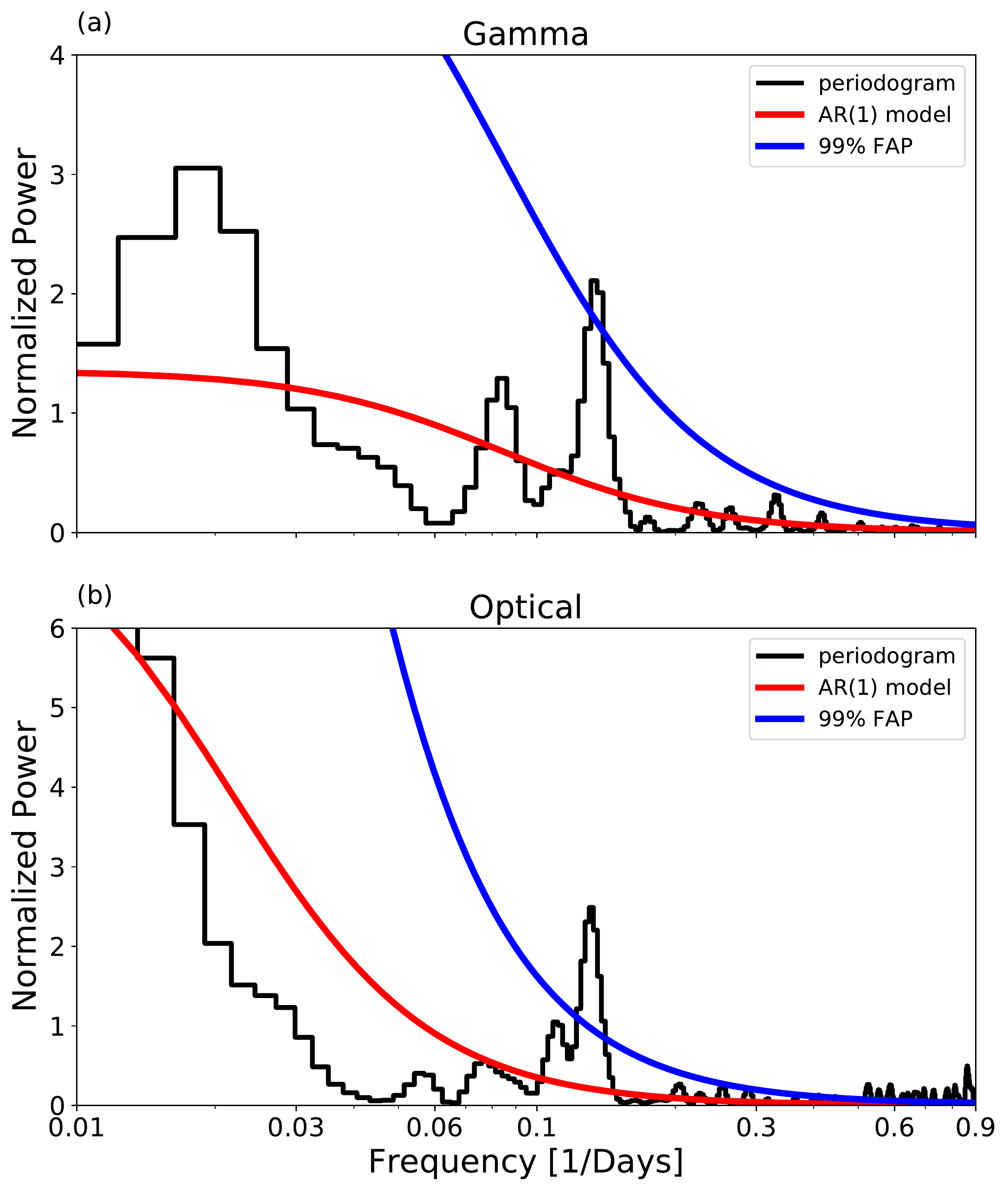}
\caption{\textbf{LSP along with the theoretical model and line of $99\%$ significance as generated by REDFIT.} \textbf{(a)} For $\gamma$-ray emissions. \textbf{(b)} For R-band
emissions.}
\label{f3}
\label{fig:fig3}
\end{figure}

\begin{figure*}[t]
\centering
\includegraphics[width=\linewidth]{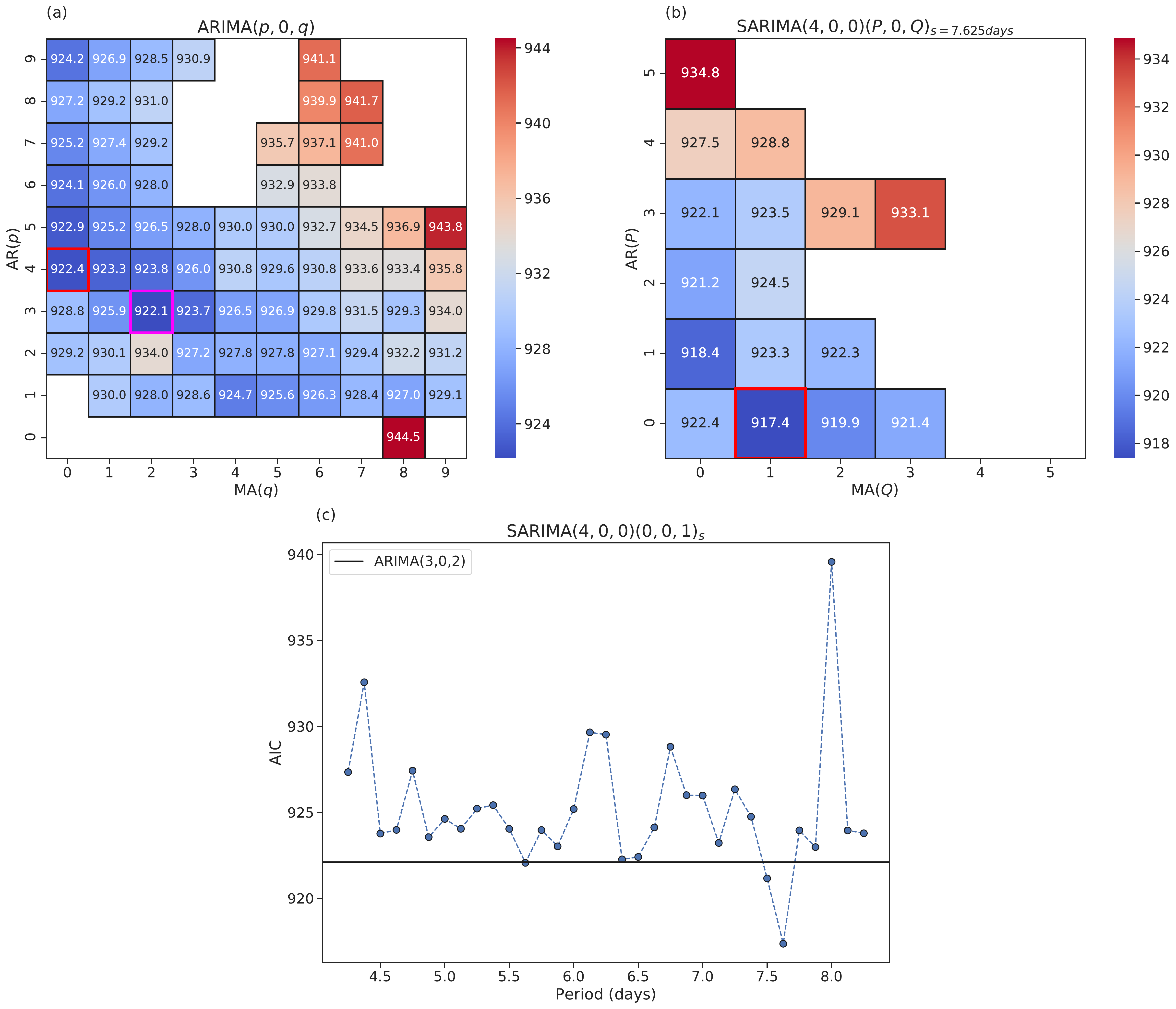}
\caption{\textbf{AIC map for the SARIMA modeling of the light curve.} \textbf{(a)} AIC map for ARIMA$(p,d,q)$ models. Though the minimum occurs at ARIMA$(3,0,2)$ (the box marked
with magenta square), the global minimum, while considering SARIMA comes at SARIMA$(4,0,0)\times(0,0,1)$ (the box marked with red square). \textbf{(b)} AIC map for
SARIMA$(4,0,0)\times(P,0,Q)$. We observe the global minimum at SARIMA$(4,0,0)\times(0,0,1)$ considering a period (season of $7.625$ days). The AIC values are provided in the boxes.
In the uncolored region, either the fit did not converge or the AIC is greater than a set threshold. This is done to improve color resolution. \textbf{(c)} AIC values for different
periods. We see a global minimum at $7.625$ days.}
\label{f4}
\end{figure*}

\section{Results}
\label{sec:4}
The three-hour binned $\gamma$-ray and {\it R}-band light curves from MJD 57710 to 57790 are presented in Fig.\ \ref{f1} along with the optimal block representation. We identify
eight significant peaks during this period. From the DCF in Fig. \ref{f1}, we also observe the $\gamma$ and optical emissions to be significantly correlated at zero lag during this
period with slightly lesser significant peaks at lags of $\approx +7$ days and $\approx -14$ days, giving us the first hint of periodicity. The WWZ map for this period shows strong
periodicity in both the wavebands, spanning from MJD 57715 to 57770, with a dominant period of $7.6^{+0.31}_{-0.17}$ in the $\gamma$-ray band and $7.6^{+0.36}_{-0.25}$ in the
optical {\it R}-band (Fig. \ref{f2}). These values were obtained by fitting the most significant peak with a log-normal function. The same apparent quasi-periods were observed in
the LSP of the two light curves in Fig. \ref{f2}. To calculate the significance, the original periodogram was fit with a bending power-law model using a maximum likelihood method.
It was confirmed (using AIC) that smooth bending power-law (AIC: 656) is a better fit for the periodograms than a regular power-law (AIC: 703). Using MC techniques from the
thousands of simulated light curves, the significance of the dominant period is $\sim 4.0\sigma$ in both the wavebands. Considering the theoretical AR1 spectrum, dominant periods
in both the wavebands show a significance of $>99\%$. The FAP for the dominant period in both the wavebands is $<10^{-5}$, although that considers an unrealistic underlying
white-noise model (Fig.\ \ref{f2}).

While modeling the light curve using seasonal ARIMA, the model with the lowest AIC is SARIMA$(4,0,0)\times(0,0,1)_{7.625 \text{d}}$ implying a periodic model better explains the
light curve (Fig.\ \ref{f4}). The SARIMA model was generated including an overall quadratic trend in the light curve during this period. The AIC values for the ARIMA models
ignoring any periodic effect are given in Fig. \ref{f4}a, while Fig. \ref{f4}b gives the AIC values including seasonal effects for $(p=4,d=0,q=0)$ with a period of $7.625$ days,
where $(P=0,D=0,Q=1)$ gives the global minimum AIC. Fig. \ref{f4}c gives the AIC values considering different periods. We find the best AIC is for a period of $7.625$ days which is
consistent with our previous observations. The best model for the light curve during this period is then
\begin{equation}
  \label{eq9}
  \mathcal{F}(t_{i}) = \sum_{j=1}^4 \theta_j\ \mathcal{F}(t_{i-j}) + \epsilon(t_i) + \epsilon(t_{i-7.625\text{d}}).
\end{equation}

It is best for ARIMA modeling if the data is evenly sampled; however, it can allow for a moderate amount of unsampled data \citep{2018FrP.....6...80F}. In the present case, we use
ARIMA only on the $\gamma$-ray light curve which has nearly perfectly ($98.4\%$) even sampling. We also tried linearly interpolating the for the small amount of unknown data and
that did not change our conclusion.

We repeated the \(\gamma\)-ray analysis with bin sizes of 12 hours and 1 day to check any role of binning in the observed results. Both of these WWZ analyses demonstrated
exactly the same periodicity with a significance of \(>4 \sigma\).

\section{Discussion}
\label{sec:5}

The results of the previous section (Sect. \ref{sec:4}) demonstrate a periodic nature in the outburst of CTA 102 during 2016-end and 2017-beginning. The dominant period of $\sim
7.6$ days is highly significant ($>4\sigma$) as is evident from the numerous statistical tests performed. It is also to be noted that the $\gamma$-ray and {\it R}-band emissions are
correlated and the dominant period is the same in both the wavebands. Also, the periodicity persists throughout the flaring period, with five QPO peaks that coincide with the
significant oberved peaks (Fig. \ref{f1}). Given that simultaneous activity was observed across this broad range of the accessible EM spectrum, it is extremely likely that the
variability originated from the jet. Since the QPO was observed for almost the entire flaring period, understanding the origin of QPO could shed light on the origin of the flare
itself.

Due to the transient nature of the quasi-periodic oscillation, we can disregard models that produce persistent QPOs such as a binary black hole system \citep{2008Natur.452..851V,
2010MNRAS.402.2087V} and persistent jet precession \citep{2000A&A...360...57R, 2004ApJ...615L...5R, 2018MNRAS.474L..81L}. Accretion-powered sources are known for exhibiting
transient QPOs which are normally believed to be related to the accretion \citep{2002A&A...387..487R}. One possible model (for QPOs with tens of days period) involve hotspots
rotating in the innermost stable circular orbit (ISCO) orbit of the SMBH \citep[e.g.,][]{1991A&A...246...21Z, 2009ApJ...690..216G}. Here the optical periodic flux modulation comes
from the circular motion of the hotspot which modulates the seed photon field for the external Compton (EC) interaction with the particles in the jet, thereby producing modulation
in the $\gamma$-ray flux \citep{2009ApJ...690..216G, 2019MNRAS.484.5785G}. This model is not favored, firstly since blazar emissions are mostly jet dominated in the optical band as
well. Also, according to that model, the $\gamma$-ray emission is Doppler boosted and thus should not have the same dominant quasi-period as the optical.

The observed blazar emission, however, is dominated by the jet, which is ultimately powered by the accretion. It is thereby prudent to focus on the mechanisms that originate in the
jet itself. QPOs related to jet are expected to be observed across the EM bands \citep[e.g.,][]{2020arXiv200301911S}. On the contrary, QPOs due to binary interaction can be
broadband like the jet or can be limited to only a few of the EM bands, depending on the interactions between the two and the mechanism responsible for the emission
\citep[e.g.,][]{2018ApJ...866...11D}.

One possible model for this jet based periodicity is magnetic reconnection events in magnetic islands inside the jet \citep{2013RAA....13..705H}. Multiple magnetic islands
(X-point reconnection) situated roughly equidistantly will have their emissions delayed by a roughly constant time, giving rise to equidistantly spaced peaks in the observed flux,
thereby mimicking rapid QPOs in standard BH systems. \cite{2018ApJ...854L..26S} observed extremely short timescale variability ($\sim 5$ min) in the source, which is smaller than
the light travel time across the central SMBH ($\sim 70$ min); hence it was attributed to the magnetic reconnection events in the magnetic islands inside the jet.

Another, purely geometrical origin of the QPO, could be a region of enhanced emission (or blob) moving helically inside the jet \citep{2015ApJ...805...91M, 2017MNRAS.465..161S}. A
region of high particle density moving in the jet can enhance the jet luminosity. Due to such helical motion, the viewing angle ($\theta_{obs}$) of the blob to line of sight varies
as \citep{2017MNRAS.465..161S, 2018NatCo...9.4599Z}
\begin{equation}
  \label{e1}
  \cos\theta_{obs}(t) = \sin \phi \ \sin \psi \cos (2\pi t/P_{obs}) + \cos \phi \ \cos\psi ,
\end{equation}
where $P_{obs}$ is the observed period, $\psi$ is the viewing angle to the jet axis from our line of sight, and $\phi$ is the pitch angle of the helical motion. This variation in
the viewing angle produces the periodic modulation in the Doppler factor ($\delta(t) = [\Gamma (1-\beta\cos \theta_{obs}(t))]^{-1}$), which gives rise to the periodic modulation of
the observed flux via $\mathcal{F}_{\nu}(t)=\delta^{3+\alpha}(t)\ \mathcal{F}'_{\nu'}$. Here $\mathcal{F}'_{\nu'}$ is the rest frame emission and $\alpha$ is the spectral index.
This model can explain the transient nature of the periodicity. The QPO begins as the blob is injected near the base of the jet and it persists until the blob dissipates. Taking
$\phi\approx 2^{\circ}$ \citep{2018NatCo...9.4599Z}, $\psi\approx 3.7^{\circ}$, with the bulk Lorentz factor, \(\Gamma = 15.5\) \citep{Hovattaetal2009} and $P_{obs} \approx 7.6$
days, the period in the rest frame of the blob would be
\begin{equation}
  \label{e2}
P=\frac{P_{obs}}{(1-\beta\cos\phi\cos\psi)}\approx 4.3\ \text{years}.
\end{equation}
The total distance traveled by the blob during one period is $D =c \beta P \cos \phi\approx 1.31$ pc. The projected distance traveled by the blob during the QPO observation is
given by $D_P=8D\sin\psi\approx 0.68$ pc. Even though it did not occur in the same epoch, the short variability timescale \(\sim 5\) min \citep{2018ApJ...854L..26S} can be
explained in this scenario by considering a blob that is smaller than the central SMBH and has a very high \(\Gamma\). This would result in very high Doppler factor (\(\delta\))
for very small \(\theta_{obs}\) and the timescales of any intrinsic fluctuations (\(\Delta t\)) in the rest frame of the blob will then be compressed to (\(\Delta t/\delta\)) in
the observed frame, producing short variability timescales. The emission from a blob moving helically inside a straight jet will give rise to periodic modulations in the observed
fluxes with nearly constant amplitude in each period. However, we do not observe this trend as the amplitudes are not constant throughout the periods. Much of this change in peak
fluxes can possibly be explained by allowing for a curvature in the jet \citep{2017Natur.552..374R, 2020arXiv200301911S}.

Of all the models discussed above, the most likely origin of the QPO, and by extension, the 2016--2017 flare, in CTA 102 is a sudden injection of a blob into the jet which then
traverses in a helical motion. This conclusion is supported by an earlier claim of observations of a helical jet structure in the source by \cite{Li_2018} in a nearby epoch where
they use a helical jet to model radio variability and VLBA component trajectories. Also, \cite{2017Natur.552..374R} explained the variability of the source using a curved jet which
is essential to explain the modulation in the peak flux in different periods. We also observe a change in the optical polarization angle from $0^{\circ}$ to $180^{\circ}$ during
the QPO period (Fig. \ref{f1}c), which is an indication of helical motions in the jet. However, one difference is that \cite{Li_2018} determined a theoretical period of $\sim 2$
years which is much longer than the one observed, but can be explained by considering a higher bulk Lorentz factor. It must be kept in mind that blazar emission mechanisms are
invariably complex, and processes like pulsational accretion flow instabilities often approximate periodic behaviors in the light curve \citep{2012MNRAS.423.3083M}.

\section{Conclusions}
\label{sec:6}

We report the detection of a significant multiwaveband ($\gamma$-ray and optical) QPO with week-like period in the flux of the blazar CTA 102 during its 2016--2017 outbursts.
Claims for the detections of multiwaveband QPOs are rare in blazar light curves \citep[to name a few]{2015ApJ...813L..41A, 2020arXiv200301911S} as are putative QPOs with week- to
month- like periods \citep{2018NatCo...9.4599Z, 2020arXiv200301911S}. We stress that several independent techniques were employed to quantify the periodicity in the light curve and
all of them produced consistent results of around $\sim 7.6$d. Nonetheless, it is necessary to keep in mind that the significance of a QPO claim can depend on the parameters (bin
sizes) and thresholds (test statistics or signal to noise ratio in each bin) adopted during the analysis. However, we found that the present $\gamma$-ray QPO detection is quite
resilient to the parameters considered (as revealed from reanalyses with larger bin sizes). From our comparison of the data with models we conclude that the most likely origin of
the flare (and the associated QPO) is a blob moving helically inside the relativistic jet.

\begin{acknowledgement}
We thank the anonymous referee for suggestions that improved the presentation of our results. This research has used data, software, and web tools of High Energy Astrophysics
Science Archive Research Center (HEASARC), maintained by NASA's Gooddard Space Flight Center. The work is also partly based on data collected by the WEBT collaboration and stored
in the WEBT archive at the Osservatorio Astrofisico di Torino -- INAF (http://www.oato.inaf.it/blazars/webt/); for questions regarding their availability, please contact the WEBT
President Massimo Villata ({\tt massimo.villata@inaf.it}). AS and VRC acknowledge support of the Department of Atomic Energy, Government of India, under project no.
12-R\&D-TFR-5.02-0200. PK acknowledges an ARIES Aryabhatta Postdoctoral Fellowship (AO/A-PDF/770).
\end{acknowledgement}

\bibliography{QPO_v2}
\bibliographystyle{aa}
\end{document}